\documentclass[prd,aps,preprint,tightenlines,showpacs,nofootinbib,superscriptaddress]{revtex4-1}
\usepackage{mathrsfs}
\usepackage{amsfonts}
\usepackage{amsmath}
\usepackage{amssymb}
\usepackage{array}
\usepackage{verbatim}
\usepackage{bm}
\usepackage{epsfig}
\usepackage{graphicx,color}
\usepackage{relsize}
\usepackage{lineno}
\usepackage{float}
\usepackage{multirow}
\RequirePackage{xspace}

\begin{document}

\begin{flushright}
MS-TP-22-53
\end{flushright}

\title{\boldmath Timelike Compton scattering in ultraperipheral $pPb$ collisions}

\author{Ya-Ping Xie}
\email{xieyaping@impcas.ac.cn}
\affiliation{Institute of Modern Physics, Chinese Academy of Sciences,
	Lanzhou 730000, China}
\affiliation{University of Chinese Academy of Sciences, Beijing 100049, China}

\author{V.~P. Gon\c{c}alves}
\email{barros@ufpel.edu.br}
\affiliation{Institut f\"ur Theoretische Physik, Westf\"alische Wilhelms-Universit\"at M\"unster,
Wilhelm-Klemm-Straße 9, D-48149 M\"unster, Germany
}
\affiliation{Institute of Modern Physics, Chinese Academy of Sciences,
  Lanzhou 730000, China}
\affiliation{Institute of Physics and Mathematics, Federal University of Pelotas, \\
  Postal Code 354,  96010-900, Pelotas, RS, Brazil}

\begin{abstract}
The study of exclusive processes in photon - induced interactions allow us to probe the generalized parton distributions (GPDs), which provide information about the 3 dimensional description of the quark and gluon content of hadrons. In this paper we investigate the timelike Compton scattering (TCS) in ultraperipheral $pPb$ collisions at the LHC. Such process is characterized by the exclusive dilepton production through the subprocess $\gamma p \rightarrow \gamma^*p \rightarrow l^+ l^- p$, with the real photon in the initial state being emitted by the nucleus. Assuming a given model for the GPDs, the TCS differential cross section is estimated as well the contribution associated to the interference between the TCS and Bethe - Heitler (BH) amplitudes. Predictions for the TCS, BH and interference contributions are presented considering the kinematical range covered by the LHC detectors. 
\end{abstract}

\maketitle

\section{Introduction}
In recent years, several analyses have indicated that the  tomography picture from the hadrons can be revealed in  exclusive processes, which can be studied in electron-hadron colliders as well in ultraperipheral collisions at RHIC and LHC (For recent reviews see, e.g. Refs. \cite{eic,EicC,lhec}). Such processes are characterized by the fact that the hadron remains intact after scattering of the photon probe and the scattering amplitude can be expressed in terms of  generalized parton distributions (GPDs) \cite{Muller:1994ses,Ji:1996ek,Radyushkin:1997ki,Collins:1998be}, which provide information about the 3 dimensional description of the quark and gluon content of hadrons \cite{Diehl:2003ny,Belitsky:2005qn}. In particular, the experimental results obtained at HERA and JLab for the deeply virtual Compton scattering (DVCS) already have provided important constraints about the GPDs, with the promising expectation that the future electron - ion colliders at BNL (EIC) \cite{eic}, China (EicC) \cite{EicC}  and CERN (LHeC) \cite{lhec} will allow us to improve our understanding of the hadronic structure in a larger kinematical range. Another example of an exclusive process is the timelike Compton scattering (TCS), which is the exclusive photoproduction of a lepton pair with large invariant mass. Although the description of this process at leading \cite{Berger:2001xd,Mueller:2012sma,Boer:2015fwa} and next - to - leading order \cite{Pire:2011st,Moutarde:2013qs} have been derived many years ago, the first measurement of TCS on proton was performed only recently by the CLAS Collaboration \cite{CLAS:2021lky}.  This analysis has demonstrated that the data, obtained in a restricted kinematical domain,  is satisfactorily described assuming the validity of the factorization theorem, which separates the hard scattering process, which can be calculated using perturbation theory, from the non-perturbative dynamics encoded in GPDs. Future measurements of TCS at the EIC and LHeC are also expected to provide a better understanding of the GPDs as well complementary constraints to those obtained by studying the DVCS process \cite{eic,EicC,lhec}. 
  
\begin{figure}
	\centering
	\includegraphics[width=5in]{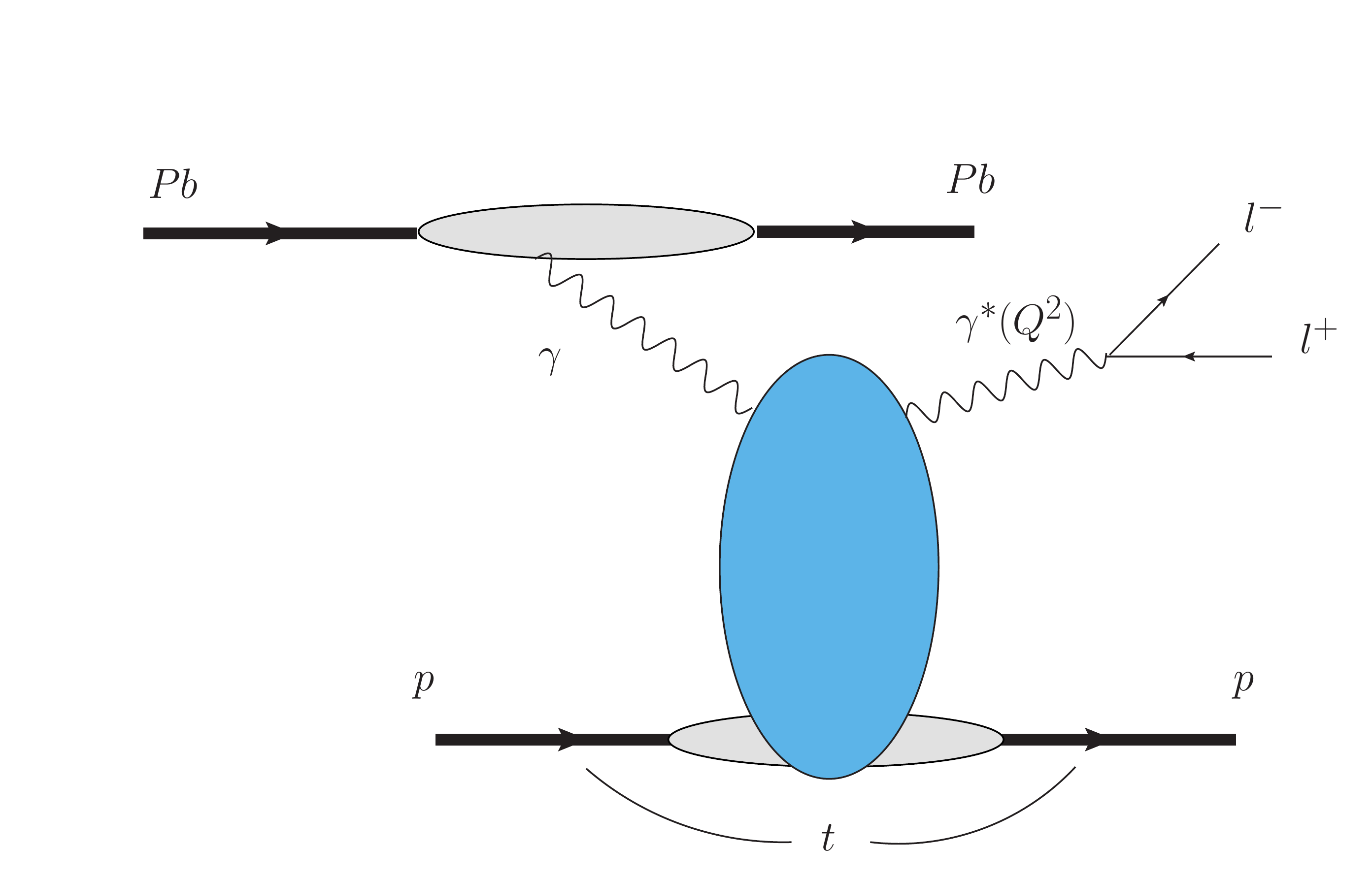}
	\caption{Timelike Compton scattering in ultraperipheral $pPb$ collisions.		 }
	\label{Fig:diagram}
\end{figure}

The study of the TCS process in proton - proton ($pp$) collisions at the LHC was proposed in Ref. \cite{Pire:2008ea} (See also Ref. \cite{Lansberg:2015kha}) considering ultraperipheral collisions (UPCs) \cite{upc}.  In these collisions, two charged hadrons (or nuclei) interact at impact parameters larger than the sum of their radii, with the hadrons acting  as a source of almost real photons, which implies that photon–hadron interactions may happen. The exploratory studies performed in Refs. \cite{Pire:2008ea,Lansberg:2015kha}  indicated that ultraperipheral collisions are a promissing way to measure the TCS process and probe the associated GPDs. Such studies strongly motivate the analysis that will be performed in this paper, where we will study the timelike Compton scattering in ultraperipheral $pPb$ collisions at the LHC energy. The representation of the process is presented in Fig. \ref{Fig:diagram}. We will focus on $pPb$ collisions due to following aspects: (a) the TCS cross section   is enhanced by a factor $Z^2$ ($Z$ is the nuclear charge) associated to the photon flux of the nucleus; (b) the UPC cross section will be dominated by photon - proton collisions, with the nucleus being the source of the photons; (c) ultraperipheral $pPb$ collisions are characterized by a low pile-up; and (d)  dileptons with small invariant mass can be measured in  these collisions. Currently, the separation of exclusive processes in high luminosity $pp$ collisions at the LHC is still a challenge due to the presence of extra $pp$ interactions per bunch crossing, usually called pile - up. The strategy to reduce the background and the impact of the pile-up is the measurement of the intact protons using the forward proton detectors (FPD) such as the ATLAS Forward Proton detector (AFP) \cite{Adamczyk:2015cjy,Tasevsky:2015xya}  and Precision Proton Spectrometer (CT-PPS) \cite{Albrow:2014lrm} that are installed symmetrically around the interaction point at a
distance of roughly 210 m from the interaction point (For a recent discussion see, e.g. Refs. \cite{Goncalves:2020saa,Martins:2022dfg}  ). However,  the current forward detectors are not able to perform the tagging of final states with invariant mass smaller than 200 GeV. As the typical values of the TCS cross section derived in Ref. \cite{Pire:2008ea} are small and strongly decrease with the invariant mass, the current experimental limitations imply that the measurement of the TCS process in $pp$ collisions is a hard task. Such difficulties are not present in $pPb$ collisions, where the exclusive processes can be separated considering that the associated events are characterized by a very low multiplicity and imposing the presence of two - rapidity gaps (regions devoid of hadronic activity) in the  final state. As in Ref. \cite{Pire:2008ea}, we will assume the validity of the equivalent photon approximation \cite{upc} to describe the ultraperipheral collisions and the TCS process will be estimated at leading order of the strong running coupling constant $\alpha_s$, which implies that the Compton amplitude is dominated by the quark handbag diagrams described in terms of the quark GPDs. The impact of the contributions proportional to gluon GPDs will be presented in a forthcoming study. It is important to emphasize that the high-energy limit of the TCS process has been estimated in the literature using alternative approaches \cite{
Schafer:2010ud,Peccini:2020jkj,Peccini:2021rbt}, as e.g. the $k_T$ - factorization and dipole approaches, which only take into account of the gluonic contribution. We also plan to compare our predictions with those derived in  Refs. \cite{
Schafer:2010ud,Peccini:2020jkj,Peccini:2021rbt} in the forthcoming study. As the nuclear photon flux is characterized by a large number of photons with small energies, one has that the typical photon - proton center - of - mass energies in ultraperipheral collisions are small, which justifies to neglect, in a first approximation, the gluonic contribution.  Our goal in this paper is to present the TCS predictions for ultraperipheral $pPb$ collisions at $\sqrt{s_{NN}} = 8.1$ TeV, as well as those associated to the Bethe - Heitler (BH) process \footnote{ The  BH process  in $pPb$ collisions was recently estimated in Ref. \cite{Linek:2022ixq}    within the $k_T$ - factorization framework considering the elastic and inelastic contributions and taking into account  the transverse momentum of the initial photons, which is an important step to improve the accuracy of the calculations.}, which contributes to the same 
final state, and for the TCS - BH interference.

 This paper is organized as follows. In the next Section, we will present a brief review of the formalism needed to derive our predictions. In particular, we will present the expressions for the TCS, BH and interference differential cross sections. Moreover, the model assumed in our analysis for the GPD will be presented. In Section \ref{sec:res} we will present our predictions for the differential distributions considering distinct values for the rapidity of the dilepton system in the final state. In addition, the results for the rapidity distribution will be shown. Finally, in Section \ref{sec:sum}, we will summarize our main results and conclusions. 
 
\section{Formalism}
In order to describe the timelike Compton scattering in ultraperipheral $pPb$ collisions we will assume the validity of the Equivalent Photon Approximation (EPA) \cite{upc}, which allow us to factorize the
 $pPb$ cross sections in terms of the photon fluxes associated to the incident particles and the photon - hadron cross sections. Although both particles act as a source of almost real photons, the nuclear photon flux is enhanced by a factor $Z^2$, which implies that the dominant contribution for the ultraperipheral $pPb$ cross sections comes from photon - proton interactions, with the nucleus acting as the photon source. Such dominance implies that the resulting rapidity distribution of the final state will be asymmetric. As the Bethe - Heitler (BH) process generates the same final state than the timelike Compton scattering, both processes contribute at the amplitude level, which implies that the differential cross section for the dilepton production in ultraperipheral $pPb$ collisions will be expressed as a sum of three contributions: $d\sigma^{total} = d\sigma^{TCS} + d\sigma^{BH} + d\sigma^{INT}$, where INT denotes the term associated to the TCS - BH interference. 
 The EPA implies that the differential cross sections for the TCS, BH and INT contributions can be expressed as follows:
\begin{eqnarray}
\left.\frac{d\sigma^i_{Pb \, p \rightarrow Pb \, l^+l^- \, p}}{dydQ^{2}dtd(\cos\theta)d\phi }\right|_{\sqrt{s_{NN}}}=n_{Pb}(\omega)\cdot \left.\frac{d\sigma^i_{\gamma \, p \rightarrow l^+l^- \,  p}}{dQ^{2}dtd(\cos\theta)d\phi }\right|_{\sqrt{s_{\gamma p}}} \,\,\,,
\label{Eq:epa}
\end{eqnarray}
where $i = $ TCS, BH or INT,  $n_{Pb}$ is the photon flux associated to the Lead nucleus and $Q^2$ is the timelike virtuality of the final state photon, which is equal to the squared invariant mass ($M^2$) of the dilepton system  that has a rapidity $y$  in the center - of - mass frame. Moreover, the polar and azimuthal angles, $\theta$ and $\phi$, of the lepton in the final state are defined as in Ref. \cite{Pire:2008ea}.    The center-of-mass energy for the photon-proton interactions is given by  $\sqrt{s_{\gamma p}} = \sqrt{4 \omega E}$, where $E = \sqrt{s_{NN}}/2$ and  $\omega$ is the photon energy which is related to the rapidity of the dilepton system by the kinematical relation $\omega = M/2 \exp (+ y)$.  The maximum photon energy can  be derived considering that the maximum possible momentum in the longitudinal direction is modified by the Lorentz factor, $\gamma_L$, due to the Lorentz contraction of the nucleus in that direction \cite{upc}. It implies $\omega_{\mbox{max}} \approx \gamma_L/R_{Pb}$ and, consequently, $\sqrt{s_{\gamma p}^{\mbox{max}}} = \sqrt{2\,\omega_{\mbox{max}}\, \sqrt{s_{NN}}}$. For $pPb$ collisions at  $\sqrt{s_{NN}} = 8.1$ TeV the LHC, one has that maximum photon -- proton center -- of -- mass energy is $\approx 1.4$ TeV \cite{upc}. However, it is important to emphasize that the photon flux is higher at smaller photon energies, i.e. it is characterized by a large number of photons with small energies and few photons with energies of the order of $\omega_{\mbox{max}}$. As a consequence, one has that, in general, $\sqrt{s_{\gamma p}} \ll \sqrt{s_{\gamma p}^{\mbox{max}}}$.   In our calculations, we will assume that  the equivalent photon spectrum generated by  the nucleus is  described by the  relativistic point-like charge model \cite{upc}, which implies that
\begin{eqnarray}
n_Z(\omega) = \frac{2Z^2\alpha_{em} }{\pi}\left\{X\,K_0(X) \cdot K_1(X) - \frac{X^2}{2}[K_1^2(X)-K_0^2(X)]\right\},
\end{eqnarray}
where $X=2 \omega R_A/\gamma_L$ and $K_0$ and $K_1$ are the modified Bessel functions.

In what follows we will review the expressions for the TCS and INT differential cross sections in photon - proton interactions, which depend on the modeling of GPDs.  The BH contribution will be estimated as in Ref. \cite{Berger:2001xd}, where the differential cross section is explicity presented. In addition, as a detailed derivation of the main formulas associated to the TCS and INT contributions was presented in Refs. \cite{Berger:2001xd,Pire:2008ea, Pire:2011st}, here we will only present the final expressions, which have been used in phenomenological studies of the TCS process at the leading order in $\alpha_s$. Our calculations will be performed using the software PARTONS \cite{Berthou:2015oaw}.

One has that the TCS contribution is given by  \cite{Berger:2001xd}
\begin{eqnarray}
\frac{d\sigma^{TCS}_{\gamma \, p \rightarrow l^+l^- \,  p}}{dQ^{2}dtd(\cos\theta)d\phi} =  \frac{\alpha^3_{em}}{8\pi s_{\gamma p}^2}\frac{1}{Q^{2}}\frac{1+\cos^2\theta}{2}
\left\{(1-\eta^2)(|\mathcal{H}_1|^2 + |\widetilde{\mathcal{H}}_1|^2) \right. \nonumber \\
\left. -2\eta^2Re(\mathcal{H}_1^*\widetilde{\mathcal{E}}_1)-\eta^2
\frac{t}{4M_p^2}|\widetilde{\mathcal{E}}_1|^2\right\}\,\,,
\label{Eq:TCS}
\end{eqnarray}
where $\eta = Q^2/(2s_{\gamma p}^2 - Q^2)$, $M_p$ is the proton mass and the Compton form factors 
$\mathcal{H}_1$, $\widetilde{\mathcal{H}}_1$,  $\mathcal{E}_1$, and $\widetilde{\mathcal{E}}_1$ are expressed in terms of the hard - scattering kernels $T_{\mathcal{H}_1,\widetilde{\mathcal{H}}_1,\mathcal{E}_1,\widetilde{\mathcal{E}}_1}^{q,g}$  and  the GPDs $H$, $\tilde{H}$, $E$ and $\tilde{E}$, defined in Ref. \cite{Diehl:2003ny} (For details see Refs. \cite{Berger:2001xd,Pire:2008ea}). At leading order, one has that $T_{\mathcal{H}_1,\widetilde{\mathcal{H}}_1,\mathcal{E}_1,\widetilde{\mathcal{E}}_1}^{g} = 0$ and the expressions for the quark sector can be found in Refs. \cite{Berger:2001xd,Pire:2008ea}. The NLO kernels are presented in Refs.  \cite{Pire:2011st,Moutarde:2013qs}.
On the other hand, the contribution associated to the interference between the TCS and BH processes can be expressed as follows \cite{Berger:2001xd,Lansberg:2015kha}
\begin{eqnarray}
\frac{d\sigma^{INT}_{\gamma \, p \rightarrow l^+l^- \,  p}}{dQ^{2}dtd(\cos\theta)d\phi}  & = & -\frac{\alpha^3_{em}}{4\pi s^2_{\gamma p}}\frac{\sqrt{t_0 - t}}{-tQ}\frac{\sqrt{1 - \eta}}{\eta} \left(\cos\phi \frac{1+\cos^2\theta}{\sin\theta}\right) \nonumber \\
& \times & Re\left[F_1(t)\mathcal{H}_1
-\eta(F_1(t) +F_2(t))\widetilde{\mathcal{H}}_1
-\frac{t}{4M_p^2}F_2(t)\mathcal{E}_1\right]\,\,,
\label{Eq:INT}
\end{eqnarray}
where $t_0 = - 4M_p^2 \eta^2/(1-\eta^2)$ and we have neglected the lepton mass and assumed that 
$s_{\gamma p},\,Q^2 \gg t, M_p^2$. Moreover, $F_1(t)$ and $F_2(t)$ are the usual Dirac and Pauli form factors, with $F_2(0)$ normalized to the anomalous magnetic moment of the proton. One has that differently from the BH and TCS contributions, which are even under the transformation $\phi \rightarrow \phi + \pi$ for integration limits symmetric about $\theta = \pi/2$, the interference term is odd due to charge conjugation. Such aspect allow us to probe the Compton amplitude through a study of $\int_0^{2\pi} d\phi \cos \phi d\sigma/d\phi$.  

The main ingredient to estimate the TCS and INT cross sections is a realistic model for the GPDs. In this exploratory study we will consider the GPDs proposed and detailed in Refs. \cite{Goloskokov:2005sd,Goloskokov:2006hr,Kroll:2012sm}, usually called Goloskokov - Kroll (GK) model, which is based on fits of meson electroproduction data. In this model, the GPDs are expressed by
\begin{equation}
F_i(x, \xi, t) = \int_1^1d\beta \int_{-1+|\beta|}^{1-|\beta|}d\alpha \, \delta(\beta + \xi \alpha -x)f_i(\beta, \alpha, t).
\end{equation}
where $F = H$, $\tilde{H}$, $E$, $\tilde{E}$ and $i$ denotes the quark flavour of the double distribution $f_i$, which are given by following expressions
\begin{eqnarray}
f_i(\beta, \alpha, t) = g_i(\beta, t)h_i(\beta) \frac{\Gamma(2n_i+2)}{2^{2n_i+1}
	\Gamma^2(n_i+1)}\frac{[(1-|\beta|)^2-\alpha^2]^{n_i}}{(1-|\beta|)^{2n_i+1}}.
\end{eqnarray}
One has that $n_i$ is set to 1 for valence quarks and 2 for sea quarks. Moreover,  $h_{sea}^q (\beta,0) =q_{sea}(|\beta|)$sign$(\beta)$ and $h_{val}^q(\beta,0) = q_{val}(|\beta|)\Theta(\beta)$, 
where $q_{sea}$ and $q_{val}$ are the usual unpolarized PDFs. Finally, the $t$-dependence of the double distributions are described by $g_i(\beta, t)$, which is assumed to have a Regge behaviour with linear trajectories, being given by
\begin{eqnarray}
g_i(\beta, t) = N e^{b_0 t}|\beta|^{-\alpha(t)}(1-\beta)^n.
\end{eqnarray}
The parameters considered in our analysis are detailed in Refs.  \cite{Goloskokov:2005sd,Goloskokov:2006hr,Kroll:2012sm}. It is important to emphasize that the GK model provides a satisfactory description of the recent TCS data measured by the CLAS Collaboration \cite{CLAS:2021lky}.

\section{Results}
\label{sec:res}
In what follows we will present our predictions for the differential cross sections associated to the Bethe - Heitler (BH), timelike Compton scattering (TCS) and TCS - BH interference (INT) considering ultraperipheral $pPb$ collisions at $\sqrt{s_{NN}} = 8.1$ TeV. It is well known that the BH contribution is much larger than the TCS one if one integrates over the full phase space of the final state \cite{Berger:2001xd,Pire:2008ea,Lansberg:2015kha}. However, as the predictions for the BH process are well determined, such a contribution can in principle be subtracted in order to access the TCS and INT contributions. Another two alternatives are (a) choose a set of cuts on the kinematical variables such that the BH and TCS contributions become similar, and/or (b) consider specific observables that are sensitive to the INT contribution, as e.g. the angular distribution of the produced leptons. In this exploratory study we will present our predictions for the $\phi$ and $t$ distributions assuming  $y = 0$ and $y = 3$, which are the rapidities covered by the central (ALICE/ATLAS/CMS) and forward (LHCb) detectors at the LHC, and a fixed value for the invariant mass of the dilepton system ($Q^2 = 5.0$ GeV$^2$). Moreover, the predictions for the rapidity distributions for fixed values of $t$ and $\phi$ and integrated over these variables will be presented. We postpone for the forthcoming study, where we will compare the LO and NLO predictions, the presentation of the results for other sets of values for $Q^2$, $\phi$ and $t$, as well the discussion about the angular asymmetries.

\begin{figure}
	\centering
	\includegraphics[width=3in]{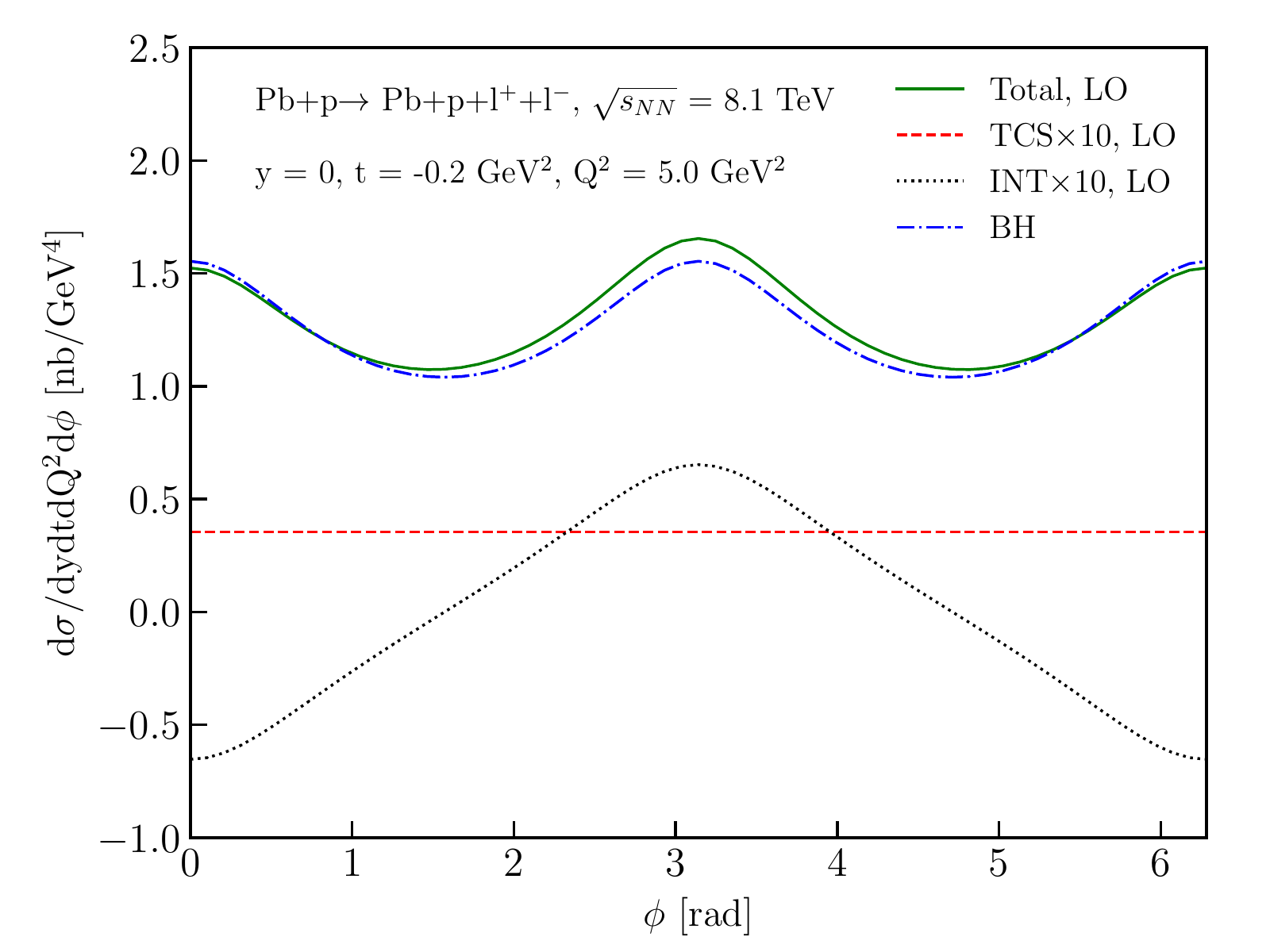}
	\includegraphics[width=3in]{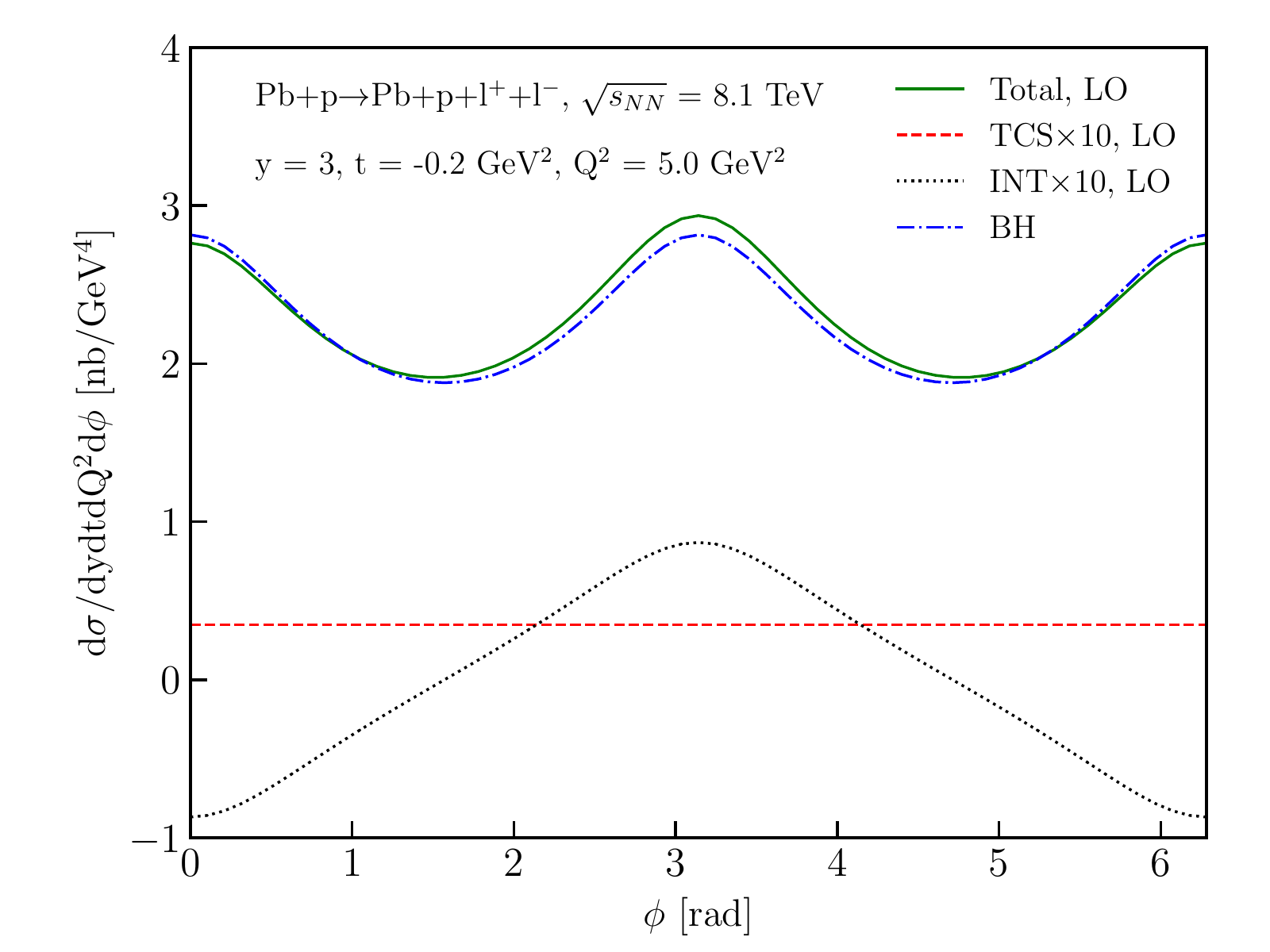}
	\caption{Azimuthal distributions of the lepton pair for  fixed values of $Q^2$ and $t$ and two values of rapidity considering ultraperipheral $pPb$ collisions at $\sqrt{s_{NN}} = 8.1$ TeV.		 }
	\label{Fig:dsdphi}
\end{figure}

In Fig. \ref{Fig:dsdphi} we present our predictions for the azimuthal dependence of the differential cross section $d\sigma/dydtdQ^2d\phi$ for fixed values of $y$, $t$ and $Q^2$, derived integrating  Eq. (\ref{Eq:epa}) over $\theta$ in the range $[\pi/4,3\pi/4]$ as in previous studies \cite{Pire:2008ea,Lansberg:2015kha}. As expected, the BH contribution dominates, with the TCS - BH interference being appreciable for $\phi \approx \pi$. Moreover, the TCS and INT contributions increase with the rapidity, which correspond to larger values of $\sqrt{s_{\gamma p}}$, in agreement with the results obtained in Ref. \cite{Pire:2008ea}. The results indicate that if the BH contribution is subtracted, the analysis of the angular distribution can be useful to probe the INT and TCS contributions.


\begin{figure}
	\centering
\includegraphics[width=3in]{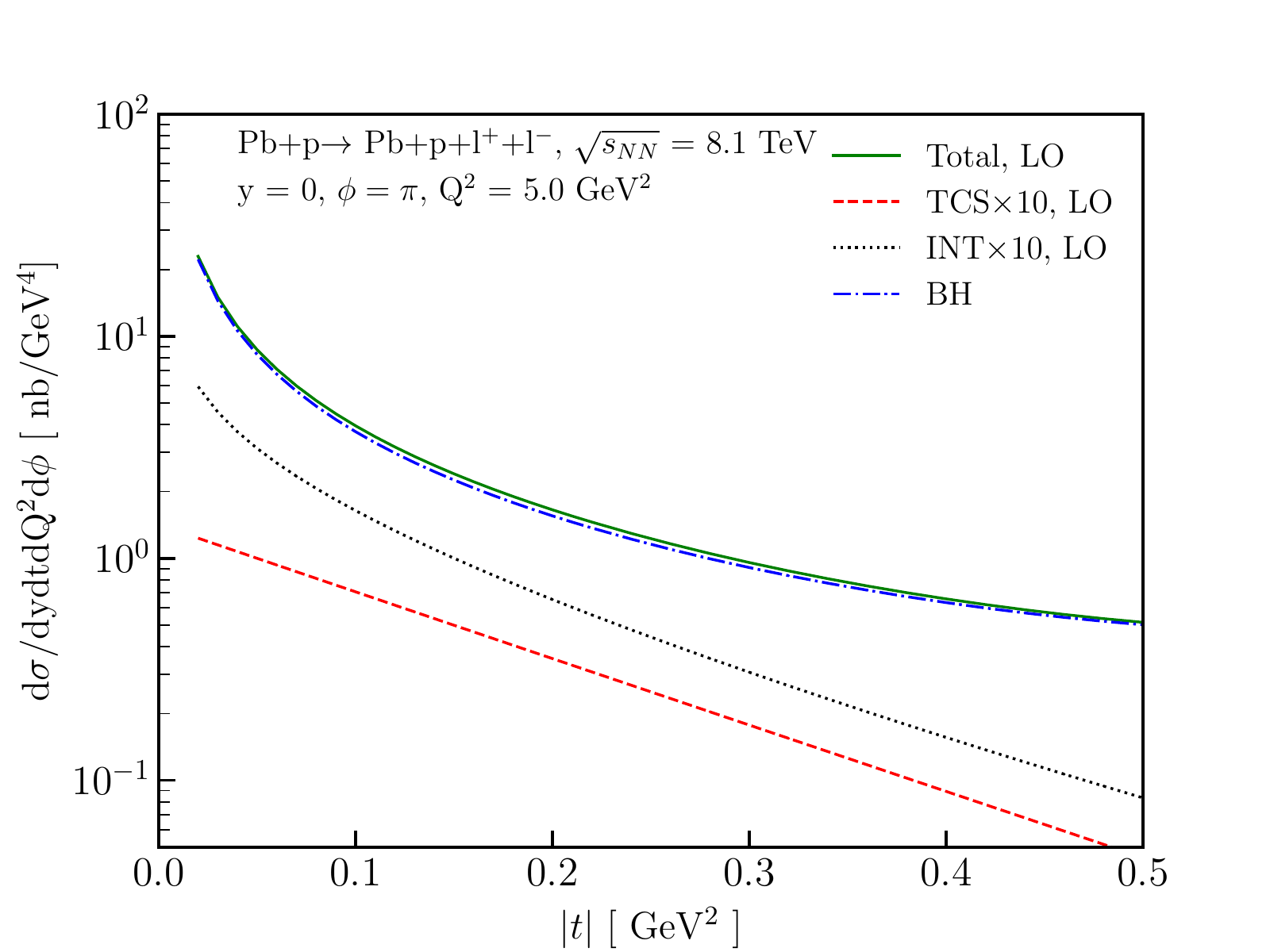}
\includegraphics[width=3in]{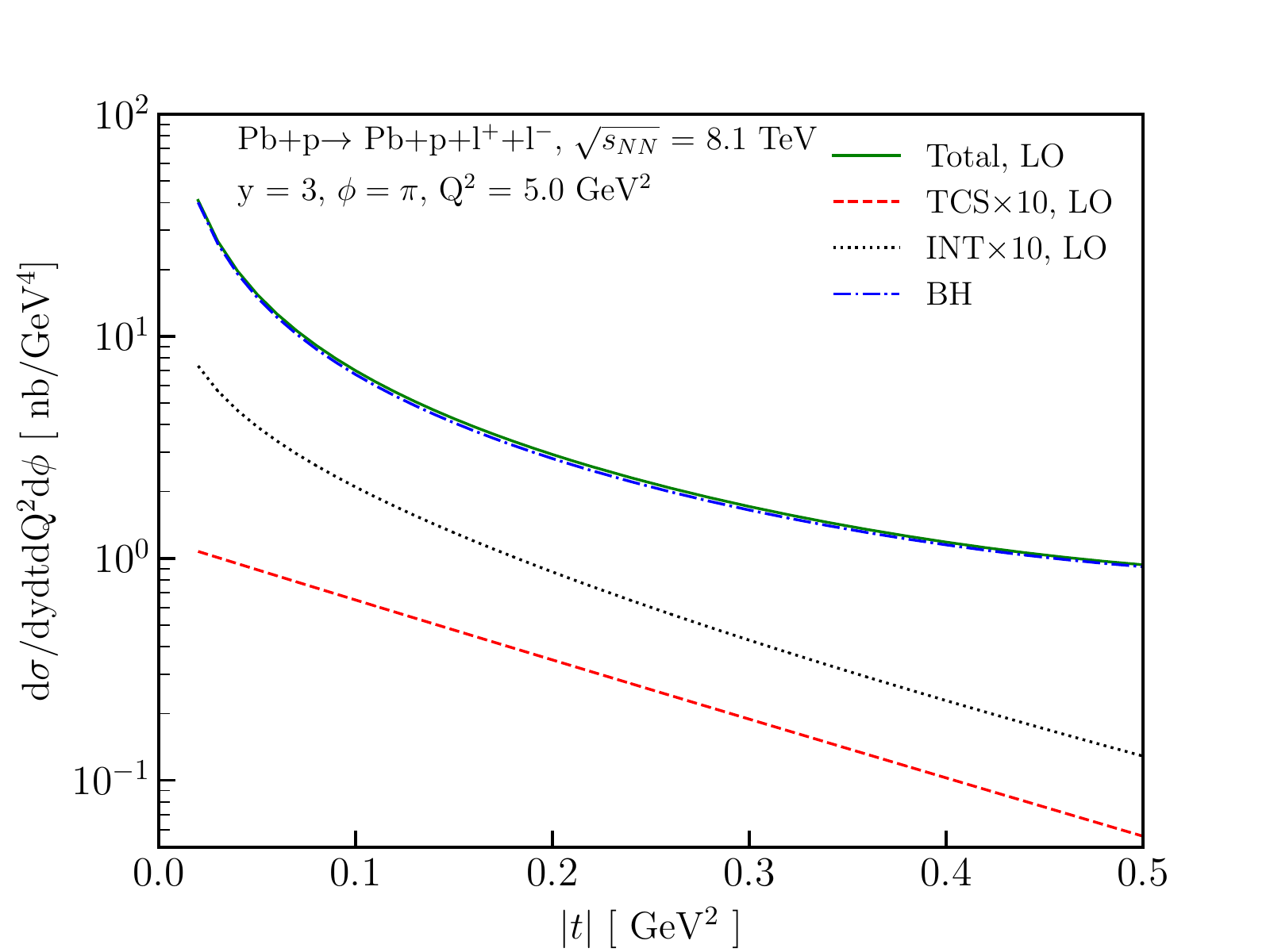}
\caption{Predictions for the $t$-distributions of the BH, TCS and INT contributions for  fixed values of $Q^2$ and $\phi$ and two values of rapidity considering ultraperipheral $pPb$ collisions at $\sqrt{s_{NN}} = 8.1$ TeV.
}
	\label{Fig:dsdt}
\end{figure}

The predictions for the $t$ - distributions of the BH, TCS and INT contributions are presented in Fig. \ref{Fig:dsdt}, assuming  fixed values for $Q^2$ and $\phi$ and two values for  the rapidity of the dilepton system. One has that the BH contribution, which is determined by the Dirac and Pauli form factors, strongly increases for $|t| \rightarrow 0$. Such increasing is also observed in the INT contribution. One has that subtracting the BH contribution and by selecting the small - $t$ events, the INT contribution will be dominant and will allow us to probe the TCS amplitude and consequently the modeling of the quark GPDs.


\begin{figure}
	\centering
	\includegraphics[width=3.2in]{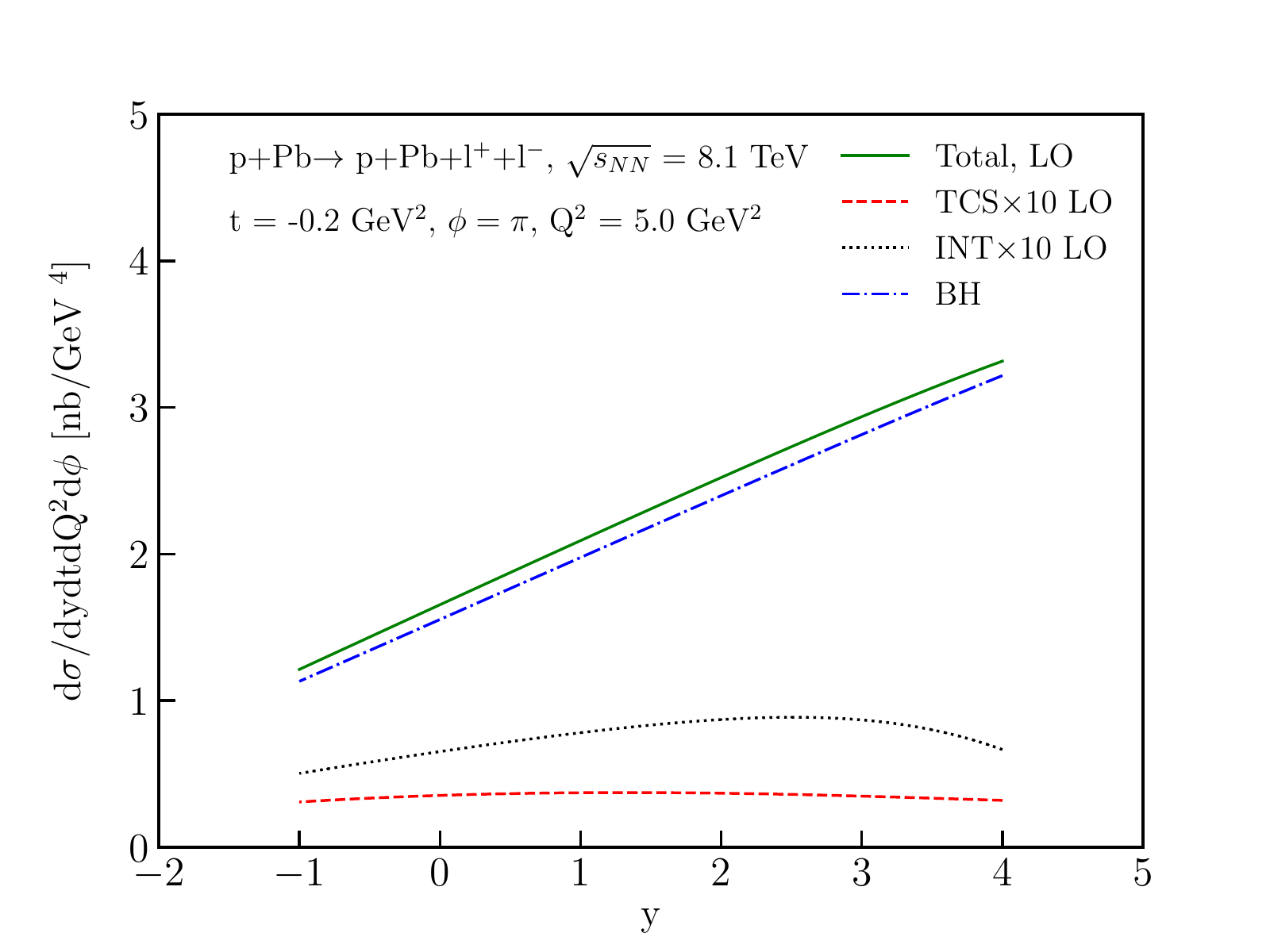}
	\includegraphics[width=3in]{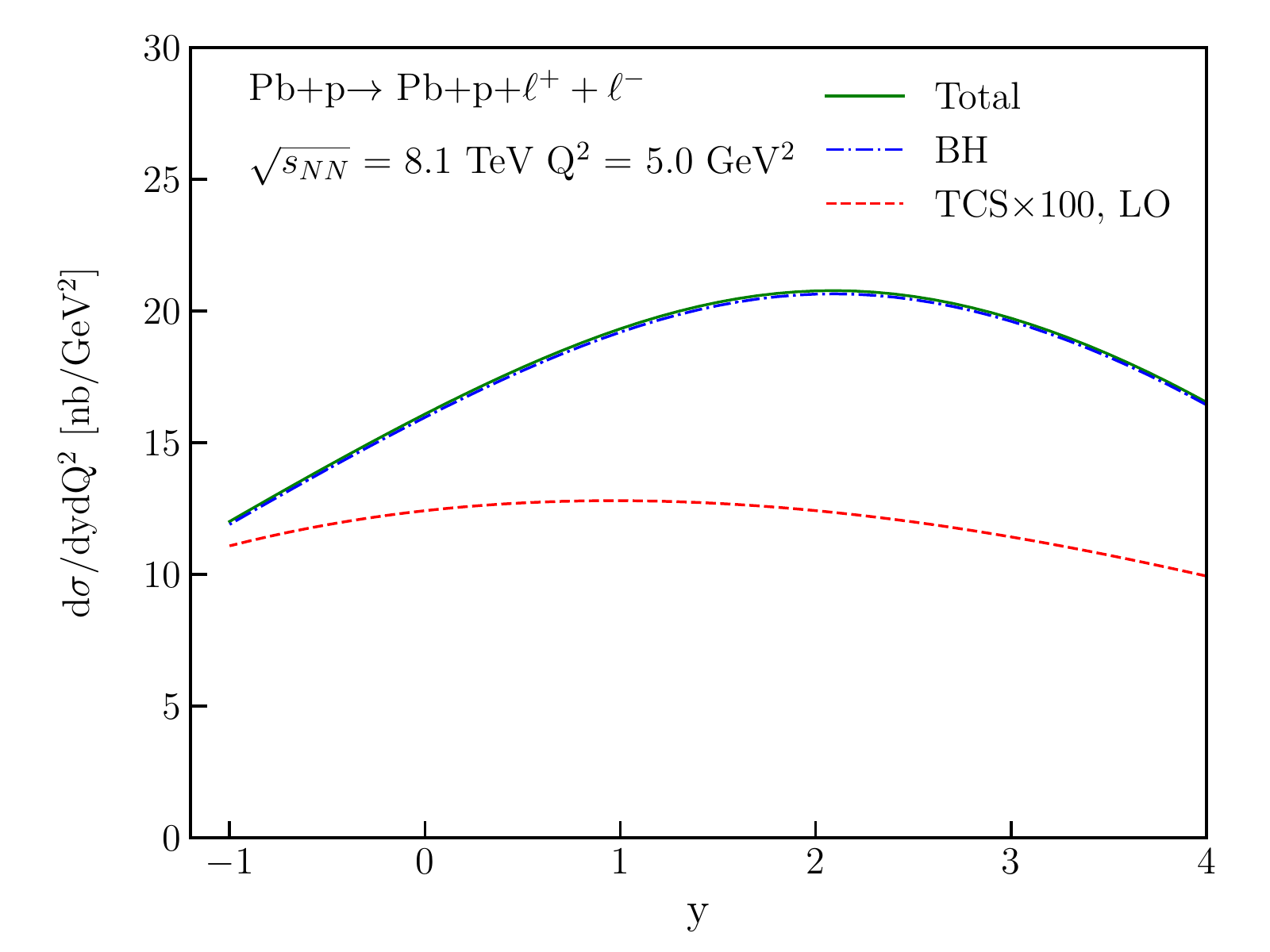}
	\caption{Predictions for the rapidity distributions  of the lepton pair produced in ultraperipheral $pPb$ collisions for  fixed values of $t$ and $\phi$ (left panel) and integrated over $t$ and $\phi$ (right panel). }
	\label{Fig:dsdy}
\end{figure}

Finally, in Fig. \ref{Fig:dsdy} we present our predictions for the rapidity distributions of the BH, TCS and INT contributions considering the dilepton production in ultraperipheral $pPb$ collisions. While in the left panel the results are presented for a fixed value of $t$ and $\phi$, the predictions in the right panel were derived integrating over $|t|$ and $\phi$ in the ranges $[0.0,0.5]$ GeV$^2$ and $[0,2\pi]$, respectively. As expected from the previous results, the BH contribution dominates over the other contributions, presenting an increasing with the rapidity when the rapidity distribution is estimated for fixed values of $t$ and $\phi$ (left panel). When the distribution is integrated over these variables (right panel), the BH contribution peaks for $y \approx 2$. Moreover, our results indicate that the analysis of the rapidity distribution for fixed values of $t$ and $\phi$ can be useful to probe the INT contribution, if the BH one is subtracted. In addition, the predictions for the integrated rapidity distribution (right panel) imply that the TCS contribution is approximately two orders of magnitude smaller than the BH one for the kinematical range and  cuts  considered.


\section{Summary}
\label{sec:sum}
One of the main goals of Particle Physics  is to improve our understanding about the quantum 3D imaging of the partons inside the protons and nuclei by measuring the parton position, momentum, and angular momentum with unprecedent precision. The quantum information of how partons are distributed inside hadrons is encoded in the
quantum phase space Wigner distributions, which include information on both generalized parton distributions (GPDs)
and transverse momentum parton distributions (TMDs). From the experimental side, the tomography picture
from the hadrons is expected to be revealed  by measurements of exclusive processes at the EIC, EicC, LHeC, RHIC and LHC. One of the promising process to probe the quark and gluon GDPs of the proton is the timelike Compton scattering, which was recently measured for the first time by the CLAS Collaboration. Motivated for this result and for the possibility of probe the TCS process in ultraperipheral collisions, in this paper we have performed an exploratory study of the timelike Compton scattering in ultraperipheral $pPb$ collisions at the LHC. The contributions of the Bethe - Heitler, TCS and interference were estimated at leading order in $\alpha_s$ and predictions for the kinematical range covered by the LHC detectors were presented assuming the GK model for the quark GPDs.  Our results indicated that the dilepton production in ultraperipheral $pPb$ collisions is dominated by the BH contribution. However, if this contribution is subtracted, our results indicate that the TCS and INT contributions are not negligible and can be probed in future experimental analysis. Such result motivates the analysis of the impact of NLO corrections, which include the contribution associated to the gluon GPDs, as well a more detailed analysis of the angular distributions and kinematical cuts on the suppression of the BH contribution. This analysis is currently being performed and will be presented in a forthcoming  study.

\section*{Acknowledgments}
One of the authors Y. P. Xie thanks useful discussions with B.~Pire and H.~Moutarde. 
The work is partially supported by the Strategic Priority Research Program of Chinese Academy of Sciences (Grant No. XDB34030301). V.P.G. was partially supported by the CAS President's International Fellowship Initiative (Grant No.  2021VMA0019) and by CNPq, CAPES, FAPERGS and  INCT-FNA (process number 
464898/2014-5).


\begin{thebibliography}{99}


\bibitem{eic}
  D.~Boer, M.~Diehl, R.~Milner, R.~Venugopalan, W.~Vogelsang, D.~Kaplan, H.~Montgomery and S.~Vigdor {\it et al.},
  arXiv:1108.1713 [nucl-th];
  A.~Accardi, J.~L.~Albacete, M.~Anselmino, N.~Armesto, E.~C.~Aschenauer, A.~Bacchetta, D.~Boer and W.~Brooks {\it et al.},
Eur.\ Phys.\ J.\ A {\bf 52}, no. 9, 268 (2016);  E.~C.~Aschenauer {\it et al.},
  Rept.\ Prog.\ Phys.\  {\bf 82}, no. 2, 024301 (2019); R.~Abdul Khalek, A.~Accardi, J.~Adam, D.~Adamiak, W.~Akers, M.~Albaladejo, A.~Al-bataineh, M.~G.~Alexeev, F.~Ameli and P.~Antonioli, \textit{et al.}
Nucl. Phys. A \textbf{1026}, 122447 (2022);
V.~Burkert, L.~Elouadrhiri, A.~Afanasev, J.~Arrington, M.~Contalbrigo, W.~Cosyn, A.~Deshpande, D.~Glazier, X.~Ji and S.~Liuti, \textit{et al.}
[arXiv:2211.15746 [nucl-ex]].



\bibitem{EicC}
D.~P.~Anderle, V.~Bertone, X.~Cao, L.~Chang, N.~Chang, G.~Chen, X.~Chen, Z.~Chen, Z.~Cui and L.~Dai, \textit{et al.}
Front. Phys. (Beijing) \textbf{16}, no.6, 64701 (2021).

\bibitem{lhec}
  J.~L.~Abelleira Fernandez {\it et al.}  [LHeC Study Group Collaboration],
  J.\ Phys.\ G {\bf 39}, 075001 (2012); P.~Agostini {\it et al.},
J. Phys. G \textbf{48}, no.11, 110501 (2021).


\bibitem{Muller:1994ses}
D.~M\"uller, D.~Robaschik, B.~Geyer, F.~M.~Dittes and J.~Ho\v{r}ej\v{s}i,
Fortsch. Phys. \textbf{42}, 101-141 (1994)

\bibitem{Ji:1996ek}
X.~D.~Ji,
Phys. Rev. Lett. \textbf{78}, 610-613 (1997)

\bibitem{Radyushkin:1997ki}
A.~V.~Radyushkin,
Phys. Rev. D \textbf{56}, 5524-5557 (1997)

\bibitem{Collins:1998be}
J.~C.~Collins and A.~Freund,
Phys. Rev. D \textbf{59}, 074009 (1999)


\bibitem{Diehl:2003ny}
M.~Diehl,
Phys. Rept. \textbf{388}, 41-277 (2003)

\bibitem{Belitsky:2005qn}
A.~V.~Belitsky and A.~V.~Radyushkin,
Phys. Rept. \textbf{418}, 1-387 (2005)









\bibitem{Berger:2001xd}
E.~R.~Berger, M.~Diehl and B.~Pire,
Eur. Phys. J. C \textbf{23}, 675-689 (2002)


\bibitem{Mueller:2012sma}
D.~Mueller, B.~Pire, L.~Szymanowski and J.~Wagner,
Phys. Rev. D \textbf{86}, 031502 (2012)


\bibitem{Boer:2015fwa}
M.~Bo\"er, M.~Guidal and M.~Vanderhaeghen,
Eur. Phys. J. A \textbf{51}, no.8, 103 (2015)



\bibitem{Pire:2011st}
B.~Pire, L.~Szymanowski and J.~Wagner,
Phys. Rev. D \textbf{83}, 034009 (2011)

\bibitem{Moutarde:2013qs}
H.~Moutarde, B.~Pire, F.~Sabatie, L.~Szymanowski and J.~Wagner,
Phys. Rev. D \textbf{87}, no.5, 054029 (2013)



\bibitem{CLAS:2021lky}
P.~Chatagnon \textit{et al.} [CLAS],
Phys. Rev. Lett. \textbf{127}, no.26, 262501 (2021)


	
\bibitem{Pire:2008ea}
B.~Pire, L.~Szymanowski and J.~Wagner,
Phys. Rev. D \textbf{79}, 014010 (2009)

\bibitem{Lansberg:2015kha}
J.~P.~Lansberg, L.~Szymanowski and J.~Wagner,
JHEP \textbf{09}, 087 (2015)


\bibitem{upc}
C. A. Bertulani and G. Baur, { Phys. Rep.} {\bf 163}, 299 (1988); F.~Krauss, M.~Greiner and G.~Soff,
  Prog.\ Part.\ Nucl.\ Phys.\  {\bf 39}, 503 (1997);
   C.~A. Bertulani, S.~R.~Klein and J.~Nystrand, Ann. Rev. Nucl. Part. Sci. {\bf 55}, 
271 (2005); V.~P.~Goncalves and M.~V.~T.~Machado,
  J.\ Phys.\ G {\bf 32}, 295 (2006);       A.~J.~Baltz {\it et al.},
  Phys.\ Rept.\  {\bf 458}, 1 (2008);       J.~G.~Contreras and J.~D.~Tapia Takaki,
  Int.\ J.\ Mod.\ Phys.\ A {\bf 30}, 1542012 (2015); 
      K.~Akiba {\it et al.} [LHC Forward Physics Working Group],
  J.\ Phys.\ G {\bf 43}, 110201 (2016); S.~R.~Klein and H.~Mantysaari,
Nature Rev. Phys. \textbf{1}, no.11, 662-674 (2019); S.~Klein and P.~Steinberg,
Ann.\ Rev.\ Nucl.\ Part.\ Sci.\  {\bf 70}, 323 (2020)






\bibitem{Adamczyk:2015cjy}
  L.~Adamczyk {\it et al.},
  CERN-LHCC-2015-009, ATLAS-TDR-024.

\bibitem{Tasevsky:2015xya}
  M.~Tasevsky [ATLAS Collaboration],
  AIP Conf.\ Proc.\  {\bf 1654},    090001 (2015).
  
    
  
\bibitem{Albrow:2014lrm}
  M.~Albrow {\it et al.} [CMS and TOTEM Collaborations],
  CERN-LHCC-2014-021, TOTEM-TDR-003, CMS-TDR-13.







\bibitem{Goncalves:2020saa}
V.~P.~Gon\c{c}alves, D.~E.~Martins, M.~S.~Rangel and M.~Tasevsky,
Phys. Rev. D \textbf{102}, no.7, 074014 (2020).


\bibitem{Martins:2022dfg}
D.~E.~Martins, M.~Tasevsky and V.~P.~Goncalves,
Phys. Rev. D \textbf{105}, no.11, 114002 (2022).


\bibitem{Schafer:2010ud}
W.~Schafer, G.~Slipek and A.~Szczurek,
Phys. Lett. B \textbf{688}, 185-191 (2010)

\bibitem{Peccini:2020jkj}
G.~M.~Peccini, L.~S.~Moriggi and M.~V.~T.~Machado,
Phys. Rev. D \textbf{102}, no.9, 094015 (2020)

\bibitem{Peccini:2021rbt}
G.~M.~Peccini, L.~S.~Moriggi and M.~V.~T.~Machado,
Phys. Rev. D \textbf{103}, no.5, 054009 (2021)


\bibitem{Linek:2022ixq}
B.~Linek, M.~\L{}uszczak, W.~Sch\"afer and A.~Szczurek,
Phys. Rev. D \textbf{106}, no.9, 094034 (2022)


\bibitem{Berthou:2015oaw}
B.~Berthou, D.~Binosi, N.~Chouika, L.~Colaneri, M.~Guidal, C.~Mezrag, H.~Moutarde, J.~Rodr\'\i{}guez-Quintero, F.~Sabati\'e and P.~Sznajder, \textit{et al.}
Eur. Phys. J. C \textbf{78}, no.6, 478 (2018)


\bibitem{Goloskokov:2005sd}
S.~V.~Goloskokov and P.~Kroll,
Eur. Phys. J. C \textbf{42}, 281-301 (2005)

\bibitem{Goloskokov:2006hr}
S.~V.~Goloskokov and P.~Kroll,
Eur. Phys. J. C \textbf{50}, 829-842 (2007)


\bibitem{Kroll:2012sm}
P.~Kroll, H.~Moutarde and F.~Sabatie,
Eur. Phys. J. C \textbf{73}, no.1, 2278 (2013)



\end{thebibliography}
\end{document}